\documentclass[sn-basic,onecolumn]{sn-jnl}
\usepackage{multirow}
\usepackage{amsmath,amssymb,amsfonts}
\usepackage{amsthm}
\usepackage{mathrsfs}
\usepackage[title]{appendix}
\usepackage{textcomp}
\usepackage{manyfoot}
\usepackage{algorithm}
\usepackage{algorithmicx}
\usepackage{algpseudocode}
\usepackage{listings}
\usepackage[table]{xcolor}
\usepackage{array,booktabs,graphicx,tabularx,longtable,makecell}
\usepackage{tikz}
\usetikzlibrary{positioning,shapes,arrows,calc,shadows}
\usepackage[edges]{forest}
\usepackage{ragged2e}
\usepackage{url}
\usepackage{placeins}

\usepackage{tikz}
\usepackage{graphicx}
\usetikzlibrary{arrows.meta, positioning, calc, fit, backgrounds, shadows}

\newcolumntype{P}[1]{>{\RaggedRight\arraybackslash}p{#1}}
\newcolumntype{Y}{>{\RaggedRight\arraybackslash}X}
\newcommand{\tcell}[1]{\makecell[tl]{#1}}
\newcolumntype{L}[1]{>{\RaggedRight\arraybackslash}p{#1}}
\newcommand{\mth}[2]{#1~\citep{#2}}
\newcommand{\MemoryThree}{Memory$^3$}
\newcommand{\HTwoO}{H$_2$O}
\newcommand{\RthreeMem}{R$^3$Mem}
\newcommand{\MemAlpha}{Mem-$\alpha$}
\newcommand{\InfinityBench}{$\infty$Bench}

\theoremstyle{thmstyleone}

\theoremstyle{thmstyletwo}

\theoremstyle{thmstylethree}

\begin{document}

\title[Learning to Remember: Advances in Memory Mechanisms for Large Language Models]{Learning to Remember: Advances in Memory Mechanisms for Large Language Models}

\author[1,2]{Chenye Han}

\author[1,2]{Hongche Yu}

\author[1]{Mengyue Wu}

\author[3]{Dell Zhang}

\author*[2]{Jixiang Luo}
\email{jixiangluo85@gmail.com}

\affil[1]{\orgname{Shanghai Jiao Tong University}, \city{Shanghai}, \country{China}}

\affil[2]{\orgname{Institute of Artificial Intelligence, China Telecom (TeleAI)}, \city{Shanghai}, \country{China}}

\affil[3]{\orgname{XSci}, \city{London}, \country{UK}}

\abstract{Memory has become a central mechanism for extending large language models (LLMs) beyond fixed context windows, stateless interaction, and static parametric knowledge. This survey organizes the literature around three complementary dimensions: the life cycle of memory, the storage substrate, and the system role that memory plays in LLM-based agents, while using cognitive-science-inspired memory concepts as a lightweight conceptual motivation rather than as a separate technical contribution or implementation family. On this basis, we distinguish three mechanism families: nonparametric memory, where information is stored outside model parameters in text, graph, multimodal, or database-like structures; parametric memory, where information is stored or accessed through activations, key-value caches, learned modules, adapter-based parameterizations, or model weights; and memory for agent evolution, where stored experiences support reinforcement learning, continual learning, procedural reuse, context engineering, and agent harness design. Evaluation is treated as a separate cross-cutting component rather than as a memory mechanism. Unlike surveys that focus primarily on retrieval-augmented generation or long-context modeling, this paper emphasizes the full memory life cycle: construction, maintenance, retrieval, consolidation, forgetting, governance, and evaluation. By aligning conceptual motivations, computational memory mechanisms, benchmarks, and deployment concerns, the survey provides a structured roadmap for designing memory-augmented LLM agents that are persistent, adaptive, auditable, and safe to update over time.}


\maketitle

\section{Introduction}\label{sec:introduction}

The rapid advancement of large language models (LLMs) has reshaped natural language processing and artificial intelligence. Scaling has improved reasoning, comprehension, and generation~\citep{brown2020languagemodelsfewshotlearners,touvron2023llamaopenefficientfoundation}, yet conventional LLMs remain constrained by fixed context windows, static parameters, and weak mechanisms for preserving state across interactions. These limitations make it difficult for models to maintain continuity, accumulate experience, adapt to changing user needs, or reason over long-horizon tasks~\citep{bubeck2023sparksartificialgeneralintelligence}. Memory mechanisms have therefore become a key research direction for transforming LLMs from stateless generators into adaptive agents.

Introducing memory into LLMs enables models to store, retrieve, and revise information across time. In dialogue and personalization, memory supports persistent user profiles, long-term preferences, and temporal consistency, as illustrated by \mth{MemoryBank}{zhong2023memorybankenhancinglargelanguage} and \mth{COMEDY}{chen2024compressimpressunleashingpotential}; in systems with explicit lifecycle operations, such as \mth{Mem0}{chhikara2025mem0}, memory can also support correction or updating of outdated facts. In knowledge-intensive settings, memory helps agents reuse prior evidence, maintain structured knowledge, and reduce repeated retrieval. In long-horizon agent systems, memory further supports planning, reflection, skill reuse, and behavioral improvement across repeated tasks.

This survey organizes LLM memory research into a lightweight conceptual-motivation layer, three computational mechanism families, and one cross-cutting evaluation component. The conceptual layer uses cognitive-science-inspired distinctions such as episodic, semantic, working, procedural, consolidated, and context-dependent memory only to clarify the vocabulary for comparing AI memory designs; it is not treated as a standalone mechanism family or the central contribution of this survey. Among computational mechanisms, \emph{nonparametric memory} stores information outside model parameters and includes text memory, graph memory, database-backed memory, and multimodal memory. \emph{Parametric memory} stores or accesses knowledge through internal states, KV caches, trainable memory modules, adapter-based parameterizations, or direct parameter updates. \emph{Memory for agent evolution} treats memory as a reusable experience substrate for reinforcement learning, continual learning, procedural reuse, context engineering, and agent harness design. Separately, \emph{evaluation} studies how these memory mechanisms are measured in plain-text, multimodal, and simulation-based environments. This distinction avoids treating cognitive analogies or benchmarks as additional storage substrates while still making both conceptual grounding and evaluation central to the survey.

The central contribution of this survey is a \emph{conceptual-motivation--lifecycle--substrate--role--evaluation alignment framework} for comparing LLM memory research. Instead of grouping systems only by implementation type, this framework aligns five questions: which conceptual memory distinctions motivate the design without themselves constituting a memory mechanism, what lifecycle operations a memory mechanism supports, which substrate realizes the memory, what role the memory plays in an LLM-based system, and how the claimed benefit is evaluated. On this basis, the contributions of this survey are threefold. (1) We provide a consolidated map of LLM memory systems that uses cognitive-science-inspired concepts only as a compact vocabulary for motivating three computational memory-mechanism families and the evaluation resources used to test them. (2) We analyze representative methods through a shared set of lifecycle questions: what is written, where it is stored, how it is retrieved, how it is updated or forgotten, how persistent it is, and how it is governed. (3) We synthesize benchmark trends and future challenges, highlighting where existing systems still struggle with reliable writing, temporal updates, privacy, deletion, hallucination control, and long-horizon utility. This alignment-oriented framing makes the relationships among conceptual motivation, memory substrates, system roles, evaluation protocols, and deployment risks explicit, thereby providing a common basis for comparing otherwise heterogeneous memory-augmented LLM systems.

\subsection{Scope and Literature Selection}\label{subsec:scope_selection}

This survey focuses on methods in which memory is an explicit design object rather than a generic byproduct of long prompts or model scale. The literature was selected by combining keyword-driven search with backward and forward citation tracing. Search terms included combinations of \emph{LLM memory}, \emph{agent memory}, \emph{long-term memory}, \emph{episodic memory}, \emph{semantic memory}, \emph{procedural memory}, \emph{graph memory}, \emph{multimodal memory}, \emph{KV cache}, \emph{model editing}, \emph{memory benchmark}, and \emph{self-evolving agents}. The search covered peer-reviewed venues, preprints, benchmark papers, benchmark repositories, system documentation, and project pages when they introduced a concrete memory architecture or operational interface.

We include systems that specify how memory is written, stored, retrieved, updated, compressed, forgotten, governed, or evaluated. The reviewed literature covers external memory stores, graph-based memory, multimodal memory, KV-cache and activation memory, weight- or module-level memory, self-evolving agents, cognitive-science memory analogies, and benchmarks that test memory persistence or long-horizon behavior. A method is included when memory is central to the mechanism or evaluation claim, not merely an incidental implementation detail. We exclude methods that only perform one-off retrieval, prompt compression, or long-context inference unless they directly address memory persistence, lifecycle management, update behavior, or memory-guided future action. This scope makes the paper a mechanism-centered survey rather than a general survey of RAG, long-context modeling, or agent frameworks.

\subsection{Memory, RAG, and Long Context}\label{subsec:memory_rag_long_context}

Memory overlaps with retrieval-augmented generation (RAG) and long-context modeling, but it is not identical to either. RAG is primarily retrieval-centered: the original RAG formulation retrieves Wikipedia passages from a dense vector index with a neural retriever, while later RAG-style systems may use sparse, dense, or hybrid retrieval~\citep{lewis2021retrievalaugmentedgenerationknowledgeintensivenlp,karpukhin2020densepassageretrievalopendomain}. Long-context modeling is capacity-centered: the system attempts to process more tokens within a single inference context. Memory is lifecycle-centered: it requires decisions about what to store, how to represent it, when to update it, how to resolve conflicts, how to delete it, and how to evaluate whether it improves future behavior. This distinction is important because a system may have a long context window or a retriever without having reliable long-term memory.

The remainder of this paper is organized as follows. Section~\ref{sec:overview} introduces the organizing framework and memory life cycle. Section~\ref{sec:cognitive_foundations} briefly explains the cognitive-science-inspired vocabulary used only as conceptual motivation. Section~\ref{sec:nonparametric_memory} reviews nonparametric memory. Section~\ref{sec:parametric_memory} discusses parametric memory. Section~\ref{sec:agent_evolution} surveys memory for agent evolution. Section~\ref{sec:benchmarks} reviews benchmarks and evaluation environments. Section~\ref{sec:challenges} discusses open challenges, followed by the conclusion.

\section{Overview}\label{sec:overview}

Research on memory in LLMs has expanded from early long-context retrieval and dialogue summarization to an ecosystem of memory stores, memory operating systems, graph memories, multimodal memories, parametric memories, and self-evolving agents. To organize this literature, we use cognitive-science-inspired memory concepts only as \emph{conceptual motivation}, and compare computational methods along three complementary dimensions: the \emph{life cycle} of memory, the \emph{form or substrate} of memory, and the \emph{system role} played by memory.

Based on these dimensions, Figure~\ref{fig:memory_taxonomy_adaptive} presents the taxonomy used throughout this survey. The taxonomy now separates a compact \emph{conceptual motivation} layer from the computational mechanism families. Cognitive-science-inspired concepts motivate how memory can be decomposed and evaluated, but the computational analysis centers on three families: nonparametric memory, parametric memory, and memory for agent evolution. Evaluation is shown alongside these families as a cross-cutting component because benchmarks determine whether a proposed memory mechanism actually improves persistence, update behavior, grounding, and long-horizon utility. Table~\ref{tab:overview_taxonomy_map} further summarizes the substrate, system role, and reviewed subcategories of each survey component, while later classification tables provide representative systems and citations.

\subsection{Memory Life Cycle}\label{subsec:memory_lifecycle}

\begin{figure}[htbp]
\centering
\includegraphics[width=\textwidth]{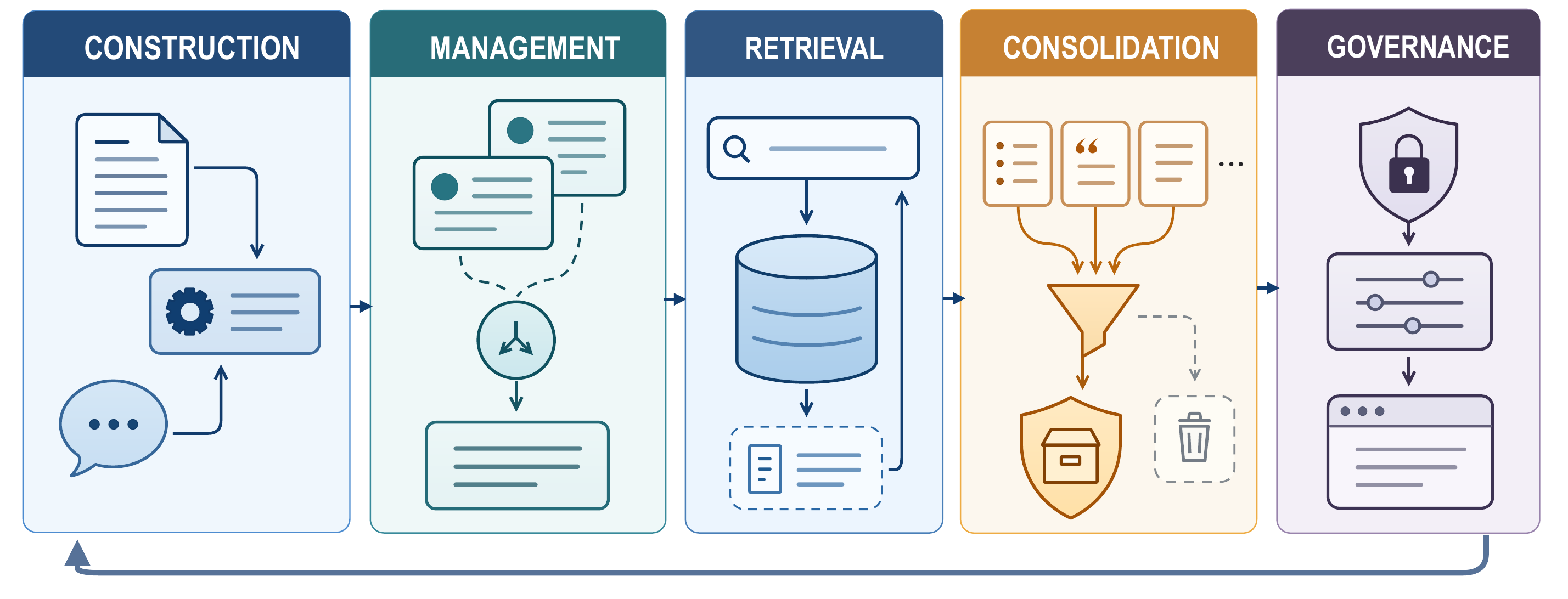}
\caption{LLM memory life cycle. Memory is constructed from source documents, conversations, perceptual streams, or model states; maintained through deduplication and conflict resolution; retrieved into the active context; consolidated or forgotten over time; and governed through access control, privacy filtering, and auditability.}
\label{fig:memory_lifecycle_overview}
\end{figure}

Although implementations differ, most memory systems can be analyzed through five recurring operations. Figure~\ref{fig:memory_lifecycle_overview} summarizes these operations as a lifecycle from memory construction to governed reuse in the LLM context.
\begin{itemize}
    \item \textbf{Construction} transforms raw interactions, documents, visual streams, or model states into memory units. Text systems summarize conversations into facts or episodes~\citep{zhong2023memorybankenhancinglargelanguage,chhikara2025mem0}; graph systems extract entities, relations, and events~\citep{gutierrez2025hipporagneurobiologicallyinspiredlongterm,rasmussen2025zep}; multimodal systems segment long videos or screen streams into captioned events~\citep{lin2025hippommhippocampalinspiredmultimodalmemory,wang2024videoagentlongformvideounderstanding}.
    \item \textbf{Management} maintains memory quality through compression, deduplication, conflict resolution, forgetting, temporal decay, permission control, versioning, and cross-session synchronization. \mth{Mem0}{chhikara2025mem0} and \mth{MemOS}{li2025memos} make memory maintenance explicit through update operations and lifecycle governance.
    \item \textbf{Retrieval} selects relevant memory and reintroduces it into the active reasoning context. Retrieval may rely on sparse retrieval such as \mth{BM25}{robertson2009probabilistic}, dense neural retrieval such as \mth{DPR}{karpukhin2020densepassageretrievalopendomain} or the original \mth{RAG}{lewis2021retrievalaugmentedgenerationknowledgeintensivenlp} retriever, later hybrid retrieval, graph traversal, temporal filters, symbolic queries, or LLM-guided search~\citep{wang2025pathpoolingtrainingfreestructure}.
    \item \textbf{Consolidation and forgetting} decide which memories should become durable, be merged into higher-level abstractions, or be discarded. This operation is increasingly important in long-running agents, where unconstrained accumulation causes noise, privacy risk, and retrieval degradation.
    \item \textbf{Evaluation and governance} measure whether memory improves downstream behavior and whether the system remains safe, private, and auditable. This dimension connects technical design to benchmark design and deployment requirements.
\end{itemize}

\subsection{Foundations, Forms, and Roles of Memory}\label{subsec:forms_roles}

The second dimension categorizes where and how memory is represented, while the conceptual-motivation layer clarifies why particular distinctions matter. \textbf{Conceptual motivation} is not a separate storage technology; instead, it provides lightweight distinctions such as episodic versus semantic memory, working versus long-term memory, procedural memory, consolidation, forgetting, context binding, and prediction-error-driven updating. \textbf{Nonparametric memory} stores information outside model parameters and is usually interpretable, updateable, and modular, but depends heavily on construction and retrieval quality. \textbf{Parametric memory} stores or accesses information through internal model states, trainable modules, KV caches, adapter-based parameterizations, or weights; it can be tightly integrated with inference but is harder to inspect and edit. \textbf{Memory for agent evolution} stores interaction traces, rewards, failures, workflows, code traces, and other experience artifacts so that agents can improve their future behavior. \textbf{Evaluation} is not a memory substrate; it is the empirical layer used to determine whether these designs produce reliable long-term behavior rather than superficial recall.

\begin{table}[htbp]
\centering
\caption{Survey-level organization of LLM memory research. The table summarizes one conceptual-motivation layer, three computational memory-mechanism families, and one cross-cutting evaluation component, including the main substrate or target, system-level role, and reviewed subcategories.}
\label{tab:overview_taxonomy_map}
\begingroup
\footnotesize
\setlength{\tabcolsep}{4pt}
\renewcommand{\arraystretch}{1.12}
\begin{tabularx}{\linewidth}{
>{\RaggedRight\arraybackslash}p{0.18\linewidth}
>{\RaggedRight\arraybackslash}p{0.23\linewidth}
>{\RaggedRight\arraybackslash}p{0.24\linewidth}
>{\RaggedRight\arraybackslash}X}
\toprule
\textbf{Survey component} &
\textbf{Where memory resides} &
\textbf{Main role in the system} &
\textbf{Reviewed subcategories} \\
\midrule
Conceptual motivation &
Cognitive-science theories and human-memory analogies rather than a computational storage substrate &
Provides the conceptual vocabulary for separating memory types, lifecycle operations, and update signals &
Episodic, semantic, working, and procedural memory; consolidation and forgetting; context binding; prediction-error updating \\
\midrule
Nonparametric memory &
External stores such as text repositories, vector databases, knowledge graphs, and multimodal archives &
Provides an explicit, inspectable, and updateable memory layer outside model weights &
Text memory; graph memory; multimodal memory for understanding and generation \\
\midrule
Parametric memory &
Internal model components such as activations, KV caches, adapter-based parameterizations, trainable modules, and model weights &
Stores or injects memory within the model's internal computation or parameters &
Activation/KV construction; KV management; weight editing; learned memory modules \\
\midrule
Memory for agent evolution &
Interaction traces, rewards, failures, workflows, code traces, and context states &
Supports behavioral improvement through experience accumulation, reflection, and skill reuse &
RL and continual learning; procedural and skill memory; context and harness engineering \\
\midrule
Evaluation &
Benchmarks built from textual histories, videos, images, app worlds, and simulated environments &
Measures memory reliability, persistence, update ability, and long-horizon utility &
Plain-text benchmarks; multimodal benchmarks; simulation environments \\
\bottomrule
\end{tabularx}
\endgroup
\end{table}

To make the comparison consistent across sections, Table~\ref{tab:crosscutting_comparison} lists the recurring analytical questions used to compare memory families. These questions shift the review from method enumeration toward lifecycle analysis.

\begin{table}[htbp]
\centering
\caption{Cross-cutting comparison dimensions for memory mechanisms. The same questions are used throughout the survey to move beyond method-by-method enumeration.}
\label{tab:crosscutting_comparison}
\footnotesize
\begin{tabularx}{\textwidth}{P{0.17\textwidth} P{0.18\textwidth} P{0.18\textwidth} P{0.18\textwidth} Y}
\toprule
\textbf{Dimension} & \textbf{Nonparametric memory} & \textbf{Parametric memory} & \textbf{Agent-evolution memory} & \textbf{Key design question} \\
\midrule
Memory unit & Text records, embeddings, triples, events, media anchors & Hidden states, KV pairs, adapter deltas, trainable vectors, edited weights & Trajectories, rewards, failures, workflows, skills, code traces & What exactly is treated as a reusable memory object? \\
\midrule
Write policy & Extraction, summarization, graph construction, database update & Cache construction, gradient update, adapter-based adaptation, model editing & Reflection, reinforcement learning, skill abstraction, harness state capture & Who decides what should be stored, and under what confidence threshold? \\
\midrule
Persistence & Often cross-session and inspectable & Ranges from transient session caches to durable parameter edits & Cross-task or lifelong, but usually tied to agent runtime & Does the memory survive beyond the current conversation or task? \\
\midrule
Update and forgetting & Explicit add, revise, merge, delete, decay, versioning & Harder to localize; may require cache eviction, re-editing, or retraining & Experience replay, policy revision, skill replacement, trace pruning & Can obsolete or harmful memory be corrected without side effects? \\
\midrule
Retrieval and use & Similarity search, graph traversal, temporal filtering, symbolic query & Attention reuse, learned lookup, internal activation or weight access & Case retrieval, procedure reuse, context folding, tool-state exposure & Is memory merely recalled, or does it causally improve later behavior? \\
\midrule
Governance & Strongest support for provenance, permissions, and deletion logs & Weakest support for auditability and deletion & Requires control over experience reuse and self-modification & How can the system prove what it remembered and remove it when needed? \\
\bottomrule
\end{tabularx}
\end{table}
% Float placement remains flexible to avoid prematurely ending text pages.

This conceptual layer is not included as a fourth computational mechanism in Table~\ref{tab:crosscutting_comparison}; instead, it defines the conceptual memory distinctions used when asking these lifecycle questions across external, internal, and agent-evolution systems.

\begin{figure}[htbp]
\centering
\raisebox{0pt}[102.6438pt][0pt]{%
\resizebox{0.96\textwidth}{102.6438pt}{%
% Editable vector source for Figure 2. Designed at final print size.
\begin{tikzpicture}[x=1pt,y=1pt,
  every node/.style={align=center,inner sep=0pt,outer sep=0pt},
  labeltext/.style={font=\sffamily\fontsize{6.2}{7.3}\selectfont,text=FigText},
  heading/.style={font=\sffamily\bfseries\fontsize{6.8}{7.4}\selectfont,text=FigHeading},
  link/.style={-{Latex[length=2.8pt,width=2.2pt]},draw=FigHeading,line width=.5pt},
  card/.style={draw=FigBorder,line width=.45pt,rounded corners=1.5pt,fill=FigFill}]
\definecolor{FigText}{HTML}{243342}
\definecolor{FigHeading}{HTML}{29465B}
\definecolor{FigBorder}{HTML}{B4C2D1}
\definecolor{FigFill}{HTML}{F1F5FA}
% Conceptual motivation remains separate from the computational families.
\draw[card,fill=white,dash pattern=on 2pt off 1.6pt] (0,68) rectangle (144,102.5);
\node[heading] at (72,95) {Conceptual motivation};
\node[labeltext] at (72,81) {Episodic, semantic, working, and\\procedural memory\\consolidation and forgetting};
\draw[card,fill=black!3] (202,68) rectangle (360,102.5);
\node[heading] at (281,95) {Memory lifecycle};
\node[labeltext] at (281,80.5) {write, store, retrieve, update\\consolidate, forget, govern};
\draw[link,dash pattern=on 2pt off 1.5pt] (144,81) -- (202,81);
\node[font=\sffamily\fontsize{5.5}{6.1}\selectfont,text=FigHeading] at (173,92) {conceptual\\vocabulary only};
% A distinct heading above the branch bus avoids label/arrow collisions.
\node[heading] at (134,60.5) {Computational memory families};
\draw[draw=FigHeading,line width=.5pt] (281,68) -- (281,51) -- (41,51);
\foreach \x in {41,134,227} {\draw[link] (\x,51) -- (\x,44);}
% Three mechanisms, followed by the cross-cutting evaluation component.
\foreach \x in {0,93,186} {\draw[card] (\x,0) rectangle +(82,44);}
\draw[card,fill=black!3] (279,0) rectangle (360,44);
\node[heading] at (41,35) {Nonparametric\\memory};
\node[labeltext] at (41,14.5) {text, graph, database\\and multimodal stores};
\node[heading] at (134,35) {Parametric\\memory};
\node[labeltext] at (134,14.5) {activation, KV-cache,\\adapter module, and\\weight memory};
\node[heading] at (227,35) {Agent-evolution\\memory};
\node[labeltext] at (227,14.5) {experience, skills,\\workflows, context,\\and harness state};
\node[heading] at (319.5,35) {Evaluation};
\node[labeltext] at (319.5,14.5) {plain-text, multimodal\\simulation, and\\long-horizon benchmarks};
\draw[link] (82,22) -- (93,22);
\draw[link] (175,22) -- (186,22);
\draw[link] (268,22) -- (279,22);
% Fix the footprint independently of node metrics to preserve pagination.
\pgfresetboundingbox
\path[use as bounding box] (-.225,-.225) rectangle (360.225,102.725);
\end{tikzpicture}%
}}
\caption{Revised taxonomy of LLM memory mechanisms. Cognitive-science-inspired concepts are used only as conceptual motivation and terminology, while the main body of the survey is organized around memory lifecycle operations, three computational memory families, and evaluation.}
\label{fig:memory_taxonomy_adaptive}
\end{figure}
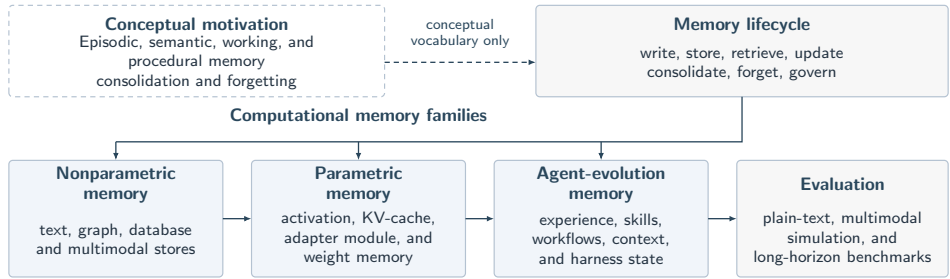

% Fig. 3: native vector redesign. All content and the caption are preserved.
% The 372pt x 189.88104pt drawing matches the original tabularx envelope.
% The figure's type sizes are final sizes: no resizebox or global changes.
\begin{figure}[htbp]
\centering
% Editable standalone artwork. Requires TikZ and xcolor.
\begingroup
\footnotesize
\definecolor{figThreeFamily}{HTML}{F1F5FA}
\definecolor{figThreeBorder}{HTML}{B4C2D1}
\definecolor{figThreeHeading}{HTML}{29465B}
\definecolor{figThreeText}{HTML}{243342}
\begin{tikzpicture}[
  x=1pt,y=1pt,baseline=-96.69052pt,
  line width=0.45pt,
  font=\sffamily\fontsize{7}{9.4}\selectfont,
  text=figThreeText,
  mechanism/.style={anchor=west,inner sep=0pt,outer sep=0pt},
  family/.style={align=left,anchor=west,inner sep=0pt,outer sep=0pt,
    text width=76pt,text=figThreeHeading,
    font=\sffamily\bfseries\fontsize{7}{9.4}\selectfont}
]
\path[use as bounding box] (0,0) rectangle (372,-189.88104);

% Equal-height bands keep the three families visually parallel.
% The shaded family column is a category label, not an extra mechanism.
\foreach \top/\bottom in {-0.225/-59.73535,-65.18535/-124.69569,-130.14569/-189.65604}{
  \path[fill=figThreeFamily,rounded corners=1.5pt]
    (0.225,\top) rectangle (371.775,\bottom);
  \path[fill=white]
    (93,\top) -- (370.275,\top)
    arc[start angle=90,end angle=0,radius=1.5pt]
    -- (371.775,\bottom+1.5)
    arc[start angle=0,end angle=-90,radius=1.5pt]
    -- (93,\bottom) -- cycle;
  \draw[figThreeBorder] (93,\top) -- (93,\bottom);
  \draw[figThreeBorder,rounded corners=1.5pt]
    (0.225,\top) rectangle (371.775,\bottom);
}

\node[family] at (9,-29.98017) {Nonparametric\\memory};
\node[family] at (9,-94.94052) {Parametric\\memory};
\node[family] at (9,-159.90087) {Memory for\\agent evolution};

% All thirteen original mechanisms are retained verbatim.
\foreach \y in {-11.18017,-20.58017,-29.98017,-39.38017,-48.78017,
  -80.84052,-90.24052,-99.64052,-109.04052,
  -145.80087,-155.20087,-164.60087,-174.00087}
  \fill[figThreeHeading] (102.5,\y) circle[radius=0.8pt];

\node[mechanism] at (108,-11.18017) {Text construction and consolidation};
\node[mechanism] at (108,-20.58017) {Text retrieval and operation control};
\node[mechanism] at (108,-29.98017) {End-to-end memory systems and databases};
\node[mechanism] at (108,-39.38017) {Triple, KG, hierarchical, and temporal graph memory};
\node[mechanism] at (108,-48.78017) {Multimodal understanding and generation memory};

\node[mechanism] at (108,-80.84052) {Activation and KV construction};
\node[mechanism] at (108,-90.24052) {KV-cache retention, eviction, paging, and compression};
\node[mechanism] at (108,-99.64052) {Weight editing and adapter-based adaptation};
\node[mechanism] at (108,-109.04052) {Learned, compressive, sparse, and decoder-style memory modules};

\node[mechanism] at (108,-145.80087) {Reinforcement and continual-learning memory};
\node[mechanism] at (108,-155.20087) {Experience replay, reflection, and credit assignment};
\node[mechanism] at (108,-164.60087) {Procedural and skill memory};
\node[mechanism] at (108,-174.00087) {Context, file, trace, and harness-state memory};
\end{tikzpicture}
\endgroup
\caption{Detailed mechanism-level taxonomy split from the original full tree. Representative systems and citations are provided in Tables~\ref{tab:nonparametric_classification}, \ref{tab:parametric_classification}, and \ref{tab:evolution_classification}, avoiding an oversized citation-heavy figure in the main text.}
\label{fig:mechanism_taxonomy_split}
\end{figure}

\section{Conceptual Motivation from Cognitive Memory}\label{sec:cognitive_foundations}

This section is deliberately brief because cognitive-science-inspired memory is used here as a source of terminology and design intuition, not as a separate technical mechanism or a primary contribution of the survey. Figure~\ref{fig:memory_taxonomy_adaptive} therefore places it outside the three computational memory families. The relevant value of cognitive memory theory is to sharpen questions that every LLM memory system must answer: whether a stored item is an event, a fact, a temporary task state, or a reusable procedure; whether it should remain local to the current session or become long-term; and whether it should be consolidated, revised, or forgotten.

Human memory research motivates this vocabulary without directly prescribing an LLM architecture. Episodic and semantic distinctions help separate event-level traces from abstracted knowledge~\citep{supp2026episodicSemanticRetrieval,supp2025humanLikeEpisodicMemory}; working-memory accounts motivate transient state management~\citep{supp2026awarenessWorkingMemory}; and studies of context binding, neural state spaces, and population codes motivate storing metadata such as time, source, scope, and retrieval conditions rather than storing decontextualized facts alone~\citep{supp2026contentContext,supp2026neuralStateSpace,supp2025neuralPopulationCodes}. These concepts are useful because many LLM-memory failures arise from collapsing different memory types into a single vector store or prompt summary. Table~\ref{tab:cognitive_design_implications} makes this role explicit by mapping cognitive concepts to LLM-memory design implications and the corresponding risks of over-analogy.

\begin{table}[htbp]
\centering
\caption{Cognitive-memory concepts as design vocabulary for LLM memory. These concepts motivate memory-system questions but should not be interpreted as evidence that LLM memory is cognitively equivalent to human memory.}
\label{tab:cognitive_design_implications}
\footnotesize
\begin{tabularx}{\textwidth}{P{0.18\textwidth} P{0.40\textwidth} Y}
\toprule
\textbf{Cognitive concept} & \textbf{LLM-memory design implication} & \textbf{Risk of over-analogy} \\
\midrule
Episodic memory & Store event-level traces with time, participants, source, and task context. & Retrieved episodes are engineered records, not human recollections. \\
\midrule
Semantic memory & Consolidate repeated observations into abstract facts, preferences, or world knowledge. & Abstraction may erase provenance, uncertainty, and conflicting versions. \\
\midrule
Working memory & Maintain transient task state, active goals, and short-lived intermediate reasoning. & A long context window or KV cache should not be confused with durable memory. \\
\midrule
Procedural memory & Preserve reusable workflows, tool-use patterns, and skills for later tasks. & Procedures may encode brittle shortcuts or unsafe habits if not validated. \\
\midrule
Consolidation and forgetting & Decide which records become durable, which are compressed, and which should decay or be deleted, especially in lifespan-oriented systems and offline recurrence settings~\citep{supp2025lifespan,supp2026sleep}. & Compression and deletion are policy choices, not direct analogues of biological consolidation or sleep. \\
\midrule
Context binding and prediction-error updating & Attach memories to source, time, scope, confidence, and surprise signals so that updates respond to conflict rather than similarity alone. & Prediction-error metaphors can hide the need for explicit verification, audit logs, and rollback. \\
\bottomrule
\end{tabularx}
\end{table}

The same caution applies to consolidation and updating. Work on prediction-error updating and hippocampal--retrosplenial communication suggests that durable memory should be affected by surprise, conflict, and context, not only by semantic similarity~\citep{supp2025predictionErrorUpdating,supp2026hippocampalRetrosplenial,supp2026dopaminergicTemporal}. In this survey, however, such findings are treated as conceptual motivation for lifecycle design rather than evidence that current LLM memories are cognitively equivalent to human memory. The core technical analysis begins with the computational substrates below: nonparametric memory, parametric memory, and memory for agent evolution.

\section{Nonparametric Memory}\label{sec:nonparametric_memory}

\begin{figure}[htbp]
\centering
\includegraphics[width=\textwidth]{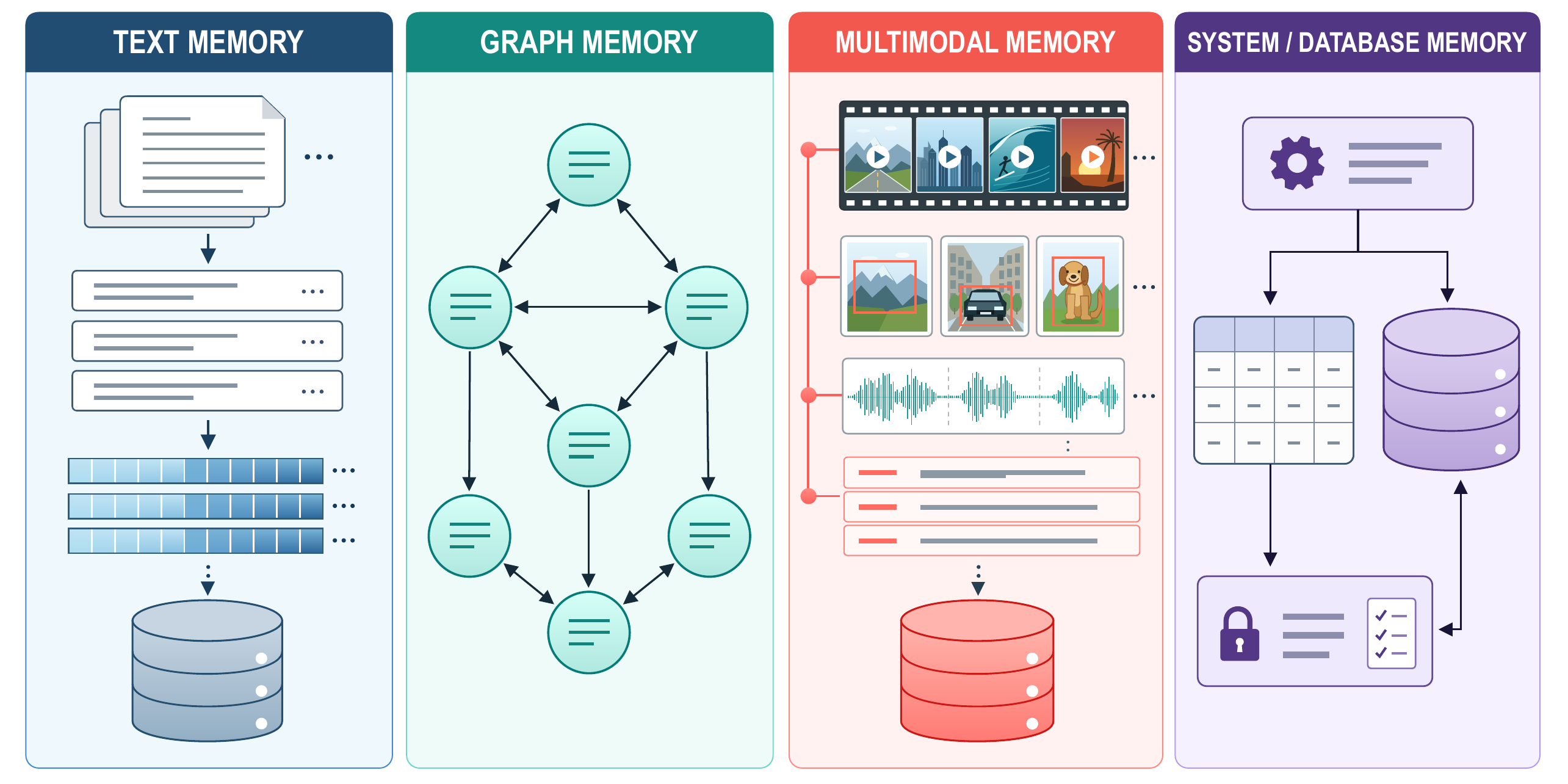}
\caption{Representative nonparametric LLM memory substrates. External memory may be represented as text chunks and embedding vectors, relational or temporal graphs, multimodal event stores, or system/database layers with controllers, metadata, indexes, permissions, and persistent storage.}
\label{fig:nonparametric_memory_substrates}
\end{figure}

Nonparametric memory stores information outside the model parameters. It is the most visible family of LLM memory methods because it supports direct inspection, deletion, correction, and cross-session persistence. In this section, we review text memory, graph memory, and multimodal memory. Text memory emphasizes compact natural-language records; graph memory emphasizes structured relations and temporal organization; multimodal memory extends memory to visual, audio, video, screen, and embodied observations. Figure~\ref{fig:nonparametric_memory_substrates} illustrates these representative external memory substrates, including database-backed memory systems that manage metadata, indexing, permissions, and persistent stores.

\subsection{Text Memory}\label{subsec:text_memory}

Text memory systems transform conversations, documents, tool traces, or observations into natural-language memory units. Their appeal is practical: text records are easy to inspect, edit, embed, and inject back into prompts. The main weakness is also structural: once events, relations, images, or tool states are collapsed into text, later retrieval may recover a plausible summary rather than the evidence needed to verify, update, or delete a memory.

\subsubsection{Memory Construction and Consolidation}\label{subsubsec:text_construction}

Construction-oriented systems differ less in the fact that they summarize history than in their write policy. \mth{MemoryBank}{zhong2023memorybankenhancinglargelanguage} and \mth{COMEDY}{chen2024compressimpressunleashingpotential} show the value of compact conversational summaries, whereas \mth{Mem0}{chhikara2025mem0}, \mth{RecMem}{dai2026recmem}, \mth{SimpleMem}{liu2026simplemem}, and \mth{LightMem}{fang2025lightmem} put more emphasis on salience, delayed consolidation, recursive compression, or staged short-/long-term updates. Personalized and temporal-dialogue systems such as \mth{context- and time-sensitive long-term memory}{supp2024temporalMemory}, \mth{Reflective Memory Management}{supp2025reflectiveMemory}, \mth{MemWeaver}{supp2025memweaver}, and \mth{Adaptive Memory Distillation}{supp2025adaptiveMemoryDistillation} further show that memory construction must decide not only how to summarize but also what deserves to become persistent, hierarchical, or reflective memory. The critical distinction is whether memory writing is eager, selective, recurrent, or sleep-like. Eager writing maximizes recall but accumulates noise and private data; delayed or selective writing reduces cost and noise but can miss rare facts that become important later. A strong text-memory system therefore needs explicit criteria for salience, conflict, source attribution, and deletion, not only a better summarizer.

\begin{table}[!htbp]
\centering
\caption{Classification of nonparametric memory methods according to this survey.}
\label{tab:nonparametric_classification}
\begingroup
\footnotesize
\setlength{\tabcolsep}{4pt}
\renewcommand{\arraystretch}{1.08}
\begin{tabularx}{\textwidth}{P{0.17\textwidth} P{0.24\textwidth} P{0.29\textwidth} Y}
\toprule
\textbf{Branch} &
\textbf{Memory representation} &
\textbf{Lifecycle emphasis} &
\textbf{Representative systems} \\
\midrule
Text memory &
Facts, events, summaries, user profiles, chunks, embeddings &
Selective writing, compression, retrieval, conflict checking, update/delete operations &
MemoryBank, Mem0, RF-Mem, Text2Mem, MemoRAG \\
\midrule
Graph memory &
Triples, entities, events, relation paths, temporal or hierarchical graphs &
Graph construction, traversal, temporal filtering, multi-hop retrieval, graph maintenance &
HippoRAG, RET-LLM, Graphiti/Zep, AriGraph, MAGMA \\
\midrule
Multimodal memory &
Frames, captions, audio transcripts, event segments, story or motion states &
Perceptual compression, timestamped retrieval, grounding, identity and temporal consistency &
DeepImageSearch, PersonaVLM, VideoAgent, VideoARM, GCAgent, StreamMeCo \\
\midrule
Memory systems and databases &
External stores, user profiles, metadata, indexes, access policies, memory resources &
End-to-end lifecycle management, scheduling, provenance, permissions, and governance &
Letta/MemGPT, MemOS, MIRIX, LangMem, OpenMemory \\
\bottomrule
\end{tabularx}
\endgroup
\end{table}

\subsubsection{Memory Retrieval and Update}\label{subsubsec:text_retrieval}

Retrieval-oriented work shifts the bottleneck from ``what to store'' to ``which stored item should affect the next answer''. Classical sparse, dense, and reranking pipelines, including BM25, \mth{DPR}{karpukhin2020densepassageretrievalopendomain}, and BERT-style rerankers~\citep{robertson2009probabilistic,gao2021rethinktrainingbertrerankers}, provide useful retrieval baselines, while \mth{Text2Mem}{wang2025text2mem}, \mth{RF-Mem}{liu2025rfmem}, \mth{MemSearch-o1}{zhang2026memsearch}, \mth{O-Mem}{wang2025mem}, \mth{MemoRAG}{supp2025memorag}, and \mth{Omne-R1}{supp2025omneR1} make retrieval more memory-specific by adding operation languages, uncertainty-aware switching, reasoning-guided search, global memory-enhanced retrieval, or memory-based multi-hop reasoning. The comparison shows that nearest-neighbor similarity is only one part of memory use: long-term agents also need temporal filters, contradiction checks, update decisions, and abstention when memory is weak or stale. Without those controls, a high-recall retriever can amplify outdated or hallucinated memories.

\subsubsection{Memory Systems and Databases}\label{subsubsec:memory_systems}

End-to-end memory platforms expose memory as an engineering interface rather than as a single algorithm. \mth{Letta/MemGPT}{packer2023memgpt} emphasizes paging between limited context and external recall or archival storage; \mth{MemOS}{li2025memos} treats memory as a schedulable resource with metadata and lifecycle operations; \mth{MIRIX}{wang2025mirix}, \mth{LangMem docs}{langchain2025langmem}, \mth{Zep}{rasmussen2025zep}, \mth{MemU repo}{memu2025}, and \mth{OpenMemory repo}{mem0ai2025openmemory} illustrate the shift toward production memory APIs, user profiles, provenance, and cross-session state. These systems are valuable because they make memory auditable and maintainable, but their claims should be separated from benchmarked algorithmic gains: documentation-driven systems often demonstrate interfaces and deployment patterns, whereas peer-reviewed methods more often isolate retrieval, update, or consolidation mechanisms.

\subsection{Graph Memory}\label{subsec:graph_memory}

Graph memory externalizes information into nodes, edges, events, and communities. Compared with flat text memory, graphs make relations and temporal links more inspectable, but they also introduce a new dependency: the memory is only as reliable as the entity extraction, relation typing, schema evolution, and graph-maintenance policy that produced it. Knowledge graphs represent entities and relations as interconnected nodes and edges~\citep{Hogan_2021,Ji_2022}; in LLM memory, this representation is attractive because many long-horizon questions require tracing who did what, when, why, and under which constraints.

Triple- and KG-based systems such as \mth{HippoRAG}{gutierrez2025hipporagneurobiologicallyinspiredlongterm}, \mth{HippoRAG2}{gutierrez2025ragmemorynonparametriccontinual}, \mth{RET-LLM}{modarressi2024retllmgeneralreadwritememory}, \mth{KGT}{sun2024knowledgegraphtuningrealtime}, \mth{Path Pooling}{wang2025pathpoolingtrainingfreestructure}, and \mth{PlugMem}{xu2025plugmem} mainly improve memory by making relational evidence retrievable. Their advantages are strongest when the task depends on multi-hop relations, personalized facts, or explainable paths. Their vulnerabilities are different from text memory: a single wrong edge can mislead several downstream paths, while aggressive graph completion can turn uncertain inferences into durable facts. This makes confidence scores, provenance, temporal validity, and edge-level deletion more important than in ordinary chunk retrieval.

Hierarchical and temporal graph methods, including \mth{Graphiti/Zep}{rasmussen2025zep}, \mth{AriGraph}{anokhin2025arigraphlearningknowledgegraph}, \mth{MemGAS}{xu2025single}, \mth{HyperMem}{yue2026hypermem}, \mth{Mnemis}{tang2026mnemis}, \mth{MAGMA}{jiang2026magma}, and \mth{TraceMem}{shu2026tracemem}, move beyond static triples by organizing memory at multiple granularities and across time. This direction is particularly relevant for agents because repeated interactions create evolving communities, goals, tasks, and causal chains rather than isolated facts. The trade-off is maintenance cost: hierarchical graph memories require decisions about when to merge nodes, split episodes, retire obsolete relations, and preserve conflicting versions. Additional work on GraphRAG-style retrieval, graph readers, synthetic KGs, hierarchical generation memory, trainable graph memory, and graph-structured personal assistants~\citep{edge2025localglobalgraphrag,li2024graphreaderbuildinggraphbasedagent,choubey2024distillsynthkgdistillingknowledgegraph,supp2025hierarchicalMemoryOrg,supp2025trainableGraphMemory,supp2025graphStructuredMemory,supp2025hiagent} is best interpreted through this lens: the key question is not whether a graph exists, but whether graph operations provide more reliable memory use than simpler text retrieval under update pressure.

\subsection{Multimodal Memory}\label{subsec:multimodal_memory}

Multimodal memory extends memory mechanisms to visual, audio, video, screen, and embodied experiences. Its central challenge is deciding what should remain as raw evidence, what should be captioned or summarized, and what should be indexed as an event. If all perceptual history is textualized, retrieval becomes cheap but grounding becomes fragile; if raw media is retained, the system preserves evidence but faces high storage, indexing, and privacy costs.

\subsubsection{Image and Visual Memory}\label{subsubsec:image_visual_memory}

Image and visual memory focuses on retaining static or sparsely changing perceptual evidence, such as image histories, screenshots, object appearances, visual preferences, and GUI observations. Compared with video memory, its main difficulty is less about dense temporal compression and more about grounding: the system must preserve object identity, spatial relations, regions, and source images so that later answers are not based only on lossy captions. \mth{DeepImageSearch}{zhang2025deepimagesearch} evaluates this setting through context-aware image retrieval over visual histories, where an agent must recover relevant images rather than merely recall text summaries. \mth{PersonaVLM}{li2025personamme} highlights personalized visual memory, storing user-specific multimodal preferences and histories so that long-term personalization can use visual evidence instead of generic user-profile text alone. Related screen and GUI memory systems, such as \mth{MGA}{supp2025mga}, treat interface observations as reusable visual state, requiring memory to preserve widgets, layouts, and interaction traces. Across these image-centered settings, the key design question is whether the memory keeps enough perceptual anchors--images, regions, screenshots, metadata, or embeddings--to support verification and correction after retrieval.

\subsubsection{Video and Temporal Multimodal Memory}\label{subsubsec:video_temporal_memory}

Video and temporal multimodal memory adds dense time, event boundaries, causal order, object trajectories, audio, and long-horizon consistency to the visual-memory problem. For understanding tasks, systems can be compared by their evidence boundary and write policy. Caption- and event-indexing methods such as \mth{Multi-RAG}{mao2025multi}, \mth{HippoMM}{lin2025hippommhippocampalinspiredmultimodalmemory}, \mth{VideoAgent}{wang2024videoagentlongformvideounderstanding}, and \mth{VideoRAG}{ren2026videorag} primarily build timestamped textual or event memories from selected frames, audio transcripts, and captions, using retrieval to avoid feeding an entire long video into the model. More explicitly agentic systems expose video memory as a tool interface: \mth{VideoAgent}{fan2024videoagentmemoryaugmented} stores temporal event descriptions and object-centric tracking states as a structured unified memory; \mth{VideoARM}{yin2026videoarm} couples an observe--think--act--memorize loop with hierarchical multimodal memory for coarse-to-fine long-video reasoning; and \mth{GCAgent}{yeo2025gcagent} builds schematic and narrative episodic memories that encode temporal and causal event relations for global-context-aware inference. For embodied settings, \mth{Embodied VideoAgent}{fan2025embodiedvideoagent} extends video memory with depth, pose, and other sensor-derived scene states, making memory persistent and updateable during interaction. Streaming systems such as \mth{HERMES}{zhang2026hermes}, \mth{EventMemAgent}{wen2026eventmemagent}, \mth{Visual Agentic Memory}{li2026visual}, and \mth{StreamMeCo}{wang2026streammeco} further shift the focus to online indexing, hierarchical retrieval, KV- or graph-based memory, and compression. These methods show that video memory is not only a question of storing more frames: it requires selective writing, recoverable visual evidence, temporal grounding, and mechanisms for controlling storage and retrieval cost.

For generation and reconstruction, memory becomes a consistency-control mechanism. \mth{MotionRAG}{yang2025motionrag}, \mth{StoryMem}{zhou2025storymem}, \mth{MemFlow}{chen2025memflow}, \mth{MagicWorld}{qiu2025magicworld}, \mth{Yume}{li2025yume}, and \mth{LoGeR}{wang2025loger} use retrieved motion, visual, story, world, or geometric state to prevent drift across long outputs. Related systems such as WorldMM, MemVerse, MGA, M3-Agent, Infinite Video Understanding, long-video memory models, OneStory, Yume-1.5, EgoLCD, Pack-and-Force Memory, Long-Term Spatial Memory, Mixture of Contexts, and Context as Memory~\citep{supp2025worldmm,supp2025memverse,supp2025mga,supp2025m3agent,supp2025infiniteVideo,supp2025episodicVideoMemory,supp2024videochatFlash,supp2024longvlm,supp2024karma,supp2026onestory,supp2025yume15,supp2025egolcd,supp2025packForce,supp2025spatialMemory,supp2025mixtureContexts,supp2025contextAsMemory} show that multimodal memory is not only about recall accuracy. It must also be evaluated by grounding fidelity, temporal coherence, identity consistency, and whether the system can trace generated content back to the perceptual or world-state memories that controlled it.

\subsection{Comparative Discussion}\label{subsec:nonparametric_discussion}

The main strength of nonparametric memory is controllability: memory items can be inspected, edited, deleted, versioned, and connected to explicit sources. However, nonparametric memory also exposes the system to retrieval errors, stale facts, noisy summaries, and inconsistent update rules. The most important design choice is not only which database or index is used, but also the memory unit and write policy. A system that stores every interaction risks noise accumulation and privacy leakage, whereas a system that stores only aggressive summaries may lose evidence needed for later contradiction resolution. Text memory is easiest to deploy but may discard structure. Graph memory supports relation-aware reasoning but depends on reliable extraction and maintenance. Multimodal memory preserves perceptual history but must balance compact captions against access to original visual or audio evidence. Table~\ref{tab:nonparametric_tradeoffs} summarizes these trade-offs.

\begin{table}[!htbp]
\centering
\caption{Trade-offs among major nonparametric memory representations.}
\label{tab:nonparametric_tradeoffs}
\footnotesize
\begin{tabularx}{\textwidth}{P{0.18\textwidth} P{0.24\textwidth} P{0.25\textwidth} Y}
\toprule
\textbf{Representation} & \textbf{Strengths} & \textbf{Limitations} & \textbf{Suitable scenarios} \\
\midrule
Text memory & Simple, editable, compatible with prompting and retrieval pipelines & Compression may lose temporal, relational, or perceptual details & Dialogue history, user profiles, personal assistants, task notes \\
\midrule
Graph memory & Preserves entities, relations, events, and multi-hop paths & Extraction noise, schema drift, and costly graph maintenance & Knowledge-intensive reasoning, temporal relation tracking, explainable retrieval \\
\midrule
Multimodal memory & Maintains links to visual, audio, video, or embodied evidence & High storage cost and difficult grounding between captions and raw media & Video agents, embodied agents, screen agents, long-form generation \\
\midrule
Database-backed memory & Supports metadata, access control, indexing, and auditability & Requires careful schema design and application integration & Long-running assistants, enterprise agents, multi-user memory systems \\
\bottomrule
\end{tabularx}
\end{table}
% Float placement remains flexible to avoid prematurely ending text pages.

\section{Parametric Memory}\label{sec:parametric_memory}

\begin{figure}[!htbp]
\centering
\includegraphics[width=0.62\textwidth]{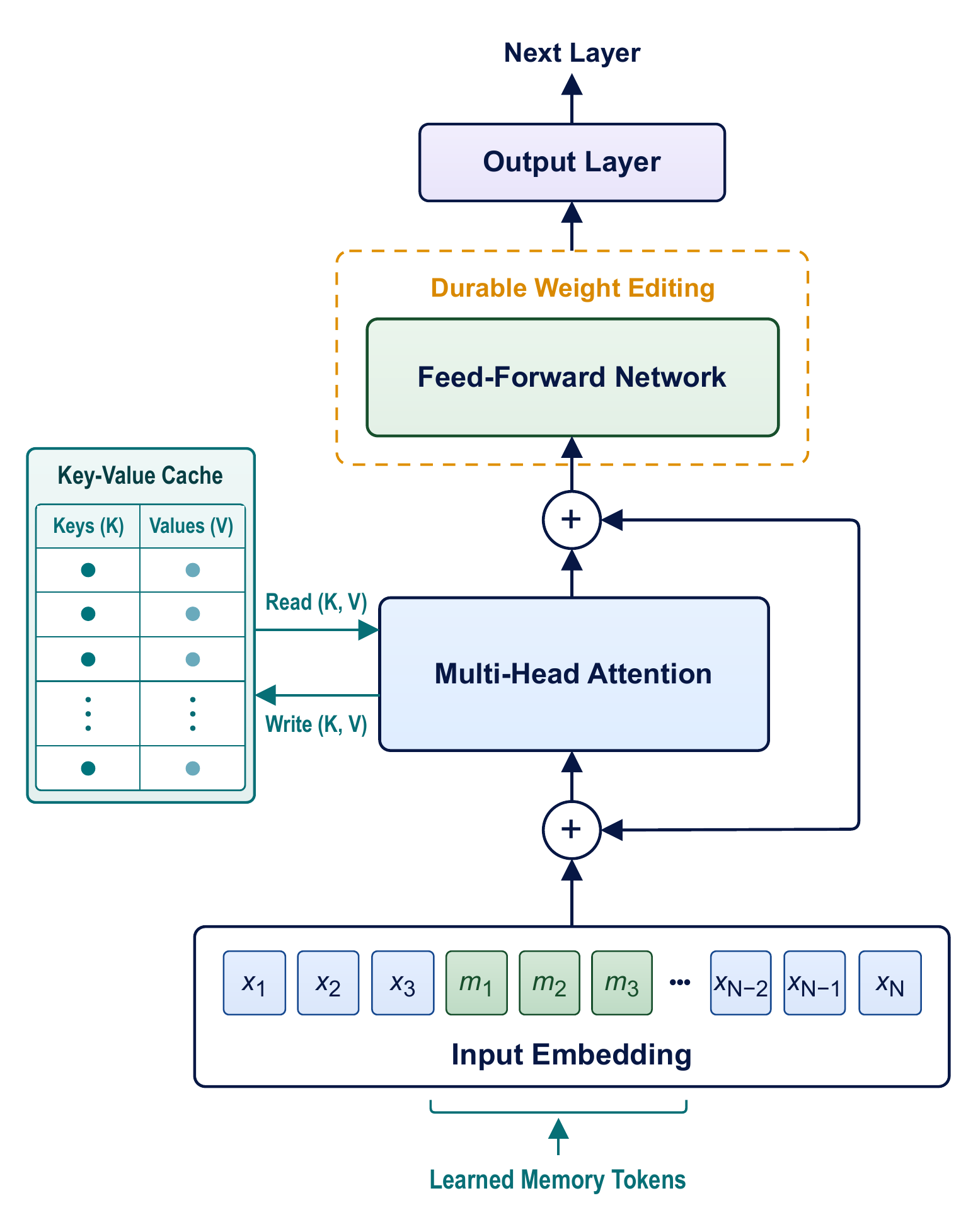}
\caption{Internal memory mechanisms inside a transformer block. KV caches provide transient session memory around attention, learned memory tokens can be introduced through the input representation, and weight editing or related parameter updates can encode more durable parametric facts in feed-forward or module-level components.}
\label{fig:internal_transformer_memory_mechanisms}
\end{figure}

Parametric memory refers to mechanisms where knowledge is stored, accessed, or updated through internal model states, trainable modules, weights, or activation caches. This category includes both transient working memory and durable internal memory. Activation states and KV caches usually function as short-term or session-level memory: they extend the effective context and improve inference efficiency, but they do not necessarily persist across independent interactions. Weight edits, trainable memory vectors, and learned modules are closer to long-term parametric memory because they can survive across sessions and alter future behavior, while adapter-based methods are best viewed as parameter-efficient adaptation carriers that become memory-relevant only when tied to a concrete update, composition, or retrieval policy. Compared with nonparametric memory, parametric memory is more tightly coupled with the model and can be efficient during inference, but it is often less transparent and harder to edit safely. Figure~\ref{fig:internal_transformer_memory_mechanisms} shows where these internal memory mechanisms can appear inside a transformer block.

% Float placement remains flexible to avoid prematurely ending text pages.

\subsection{Activation and KV-Cache Memory}\label{subsec:activation_kv_memory}

Activation and KV-cache methods are often described as memory, but their persistence is usually weaker than that of external stores or weight updates. The KV cache stores keys and values from previous tokens so the model can attend to earlier context without recomputing states; this makes it a form of working memory by default. It becomes long-term memory only when the cache, compressed states, or activation summaries are explicitly stored, retrieved, and reused across sessions.

Construction-oriented methods differ in how they extend the context available to the model. \mth{Transformer-XL}{dai2019transformerxlattentivelanguagemodels} uses segment recurrence, \mth{Memorizing Transformer}{wu2022memorizingtransformers} augments attention with stored key-value pairs, \mth{LongMem}{wang2023augmentinglanguagemodelslongterm} retrieves cached KV memories through an auxiliary network, \mth{MA-LMM}{He_2024_CVPR} stores online visual features and learned query representations, and \mth{\MemoryThree}{Hongkang_Yang_2024} frames explicit memory as a storage form beyond parameters and prompt tokens. These methods improve access to prior context, but the memory unit is often an activation state rather than an inspectable fact. As a result, they are strong for continuity and efficiency but weak for provenance, user correction, and deletion.

\begin{table}[!htbp]
\centering
\caption{Classification of parametric memory methods.}
\label{tab:parametric_classification}
\begingroup
\footnotesize
\setlength{\tabcolsep}{4pt}
\renewcommand{\arraystretch}{1.08}
\begin{tabularx}{\textwidth}{P{0.19\textwidth} P{0.25\textwidth} P{0.30\textwidth} Y}
\toprule
\textbf{Branch} &
\textbf{Internal memory object} &
\textbf{Main design focus} &
\textbf{Representative methods} \\
\midrule
Activation and KV construction &
Hidden states, attention keys and values, memory tokens &
Extend usable context through recurrence, cached attention, or explicit activation storage &
Transformer-XL, Memorizing Transformer, LongMem, MA-LMM, Memory$^3$ \\
\midrule
KV-cache management &
KV blocks, token caches, compressed attention states &
Retain, evict, merge, page, or compress caches while preserving useful context &
H$_2$O, SnapKV, PyramidKV, ChunkKV, PagedAttention, MemServe \\
\midrule
Weight editing and adapter-based adaptation &
Feed-forward weights, low-rank adapter deltas, side memories &
Update durable internal facts or task knowledge with locality and rollback control &
ROME, MEMIT, LoRA, WISE, MemoryLLM \\
\midrule
Learned and compressive memory modules &
Trainable memory vectors, virtual memory tokens, neural lookup modules, decoders, sparse experts &
Couple memory with model computation through learned lookup, compression, or test-time update &
R$^3$Mem, Engram, Memory Decoder, GradMem, STEM \\
\bottomrule
\end{tabularx}
\endgroup
\end{table}
% Float placement remains flexible to avoid prematurely ending text pages.

KV management methods address a different problem: how to keep enough context while reducing inference cost. \mth{\HTwoO}{zhang2023h2oheavyhitteroracleefficient}, \mth{SnapKV}{NEURIPS2024_28ab4182}, \mth{PyramidKV}{cai2025pyramidkvdynamickvcache}, \mth{ChunkKV}{liu2025chunkkvsemanticpreservingkvcache}, \mth{StreamingLLM}{xiao2024efficientstreaminglanguagemodels}, \mth{FastGen}{ge2024modeltellsdiscardadaptive}, \mth{MiniCache}{NEURIPS2024_fd070571}, \mth{PagedAttention}{kwon2023efficientmemorymanagementlarge}, \mth{MemServe}{hu2024memservecontextcachingdisaggregated}, \mth{PyramidInfer}{yang2024pyramidinferpyramidkvcache}, and \mth{Infinite-LLM}{lin2024infinitellmefficientllmservice} vary in whether they select heavy hitters, preserve semantic chunks, compress hierarchically, merge states, or manage cache blocks as a serving resource. The critical evaluation question is therefore not simply throughput or perplexity: a cache policy should also be tested for which memories it silently discards, whether discarded content was causally needed later, and whether compressed cache states can support update or forgetting semantics.

\subsection{Weight and Module Memory}\label{subsec:weight_module_memory}

Weight and module memory is more persistent than activation memory because it changes parameters or parameter-like states that affect future inference. This persistence is valuable when knowledge should be compact and immediately available, but it also makes errors harder to inspect and remove. The main comparison with nonparametric memory is therefore a governance trade-off: internal memory can reduce retrieval overhead, yet it weakens source traceability and precise deletion.

Model-editing methods such as \mth{ROME}{meng2023locatingeditingfactualassociations} and \mth{MEMIT}{meng2023masseditingmemorytransformer} treat selected feed-forward weights as locations for factual associations; \mth{WISE}{wang2024wiserethinkingknowledgememory} separates main and side memory to improve locality and reliability; and \mth{MemoryLLM}{wang2024memoryllmselfupdatablelargelanguage} adds a fixed-size latent memory pool inside the transformer. Adapter-based approaches such as \mth{LoRA}{hu2021loralowrankadaptationlarge} are memory-relevant when the adapter is used as a durable carrier of task, domain, user, or preference information, but standard LoRA is not by itself an online memory system because its updates are normally fixed after training. The key criterion is whether the method specifies a write, composition, retrieval, and rollback policy, not merely whether additional parameters exist.

Learned and compressive modules, including \mth{\RthreeMem}{zhang2025r3mem}, \mth{Engram}{chen2025engram}, \mth{MLP Memory}{geva2021transformerfeedforward}, \mth{Memory Decoder}{liu2025memorydecoder}, \mth{GradMem}{li2025gradmem}, \mth{MeKi}{wang2025meki}, and \mth{STEM}{zhang2025stem}, further blur the line between neural computation and memory storage. Some reduce explicit retrieval costs by making memories part of the computation graph; others trade retrieval for test-time optimization or sparse lookup. Related work on activation memory, recurrent memory, sparse memory, function-token memory, explicit working memory, infinitely updatable state, memory--reasoning disentanglement, and contextual agentic memory~\citep{supp2024infllm,supp2024titans,supp2026kvCompressibility,supp2026memoryCaching,supp2026deltaMem,supp2026msa,supp2026fastWeightPkm,supp2024ultraSparseMemory,supp2026languageMemoryModels,supp2025functionTokenMemory,supp2025nestedLearning,supp2025explicitWorkingMemory,supp2025updatableState,supp2025memo,supp2025memoryReasoning,supp2026contextualAgenticMemory} reinforces the same point: internal memory should be evaluated by persistence, locality, reversibility, interference, and privacy leakage, not only by loss or speed.

\subsection{Comparative Discussion}\label{subsec:parametric_discussion}

Parametric memory is attractive when memory must be close to model computation. KV-cache methods are useful for long-context inference and serving efficiency, but they usually preserve information only for the current or cached session unless explicitly externalized. This makes them different from persistent memory systems even though both are often discussed under the language of memory. Weight editing and learned memory modules can make memories compact and immediately accessible, but they are difficult to audit and delete. Their benefits should therefore be evaluated together with locality, reversibility, interference, and privacy risk. For safety-critical or privacy-sensitive deployments, parametric memory should be paired with external provenance records, edit logs, and evaluation protocols that test whether an internal update changes only the intended behavior rather than silently spreading to unrelated contexts.

\section{Memory for Agent Evolution}\label{sec:agent_evolution}

\begin{figure}[!htbp]
\centering
\includegraphics[width=0.62\textwidth]{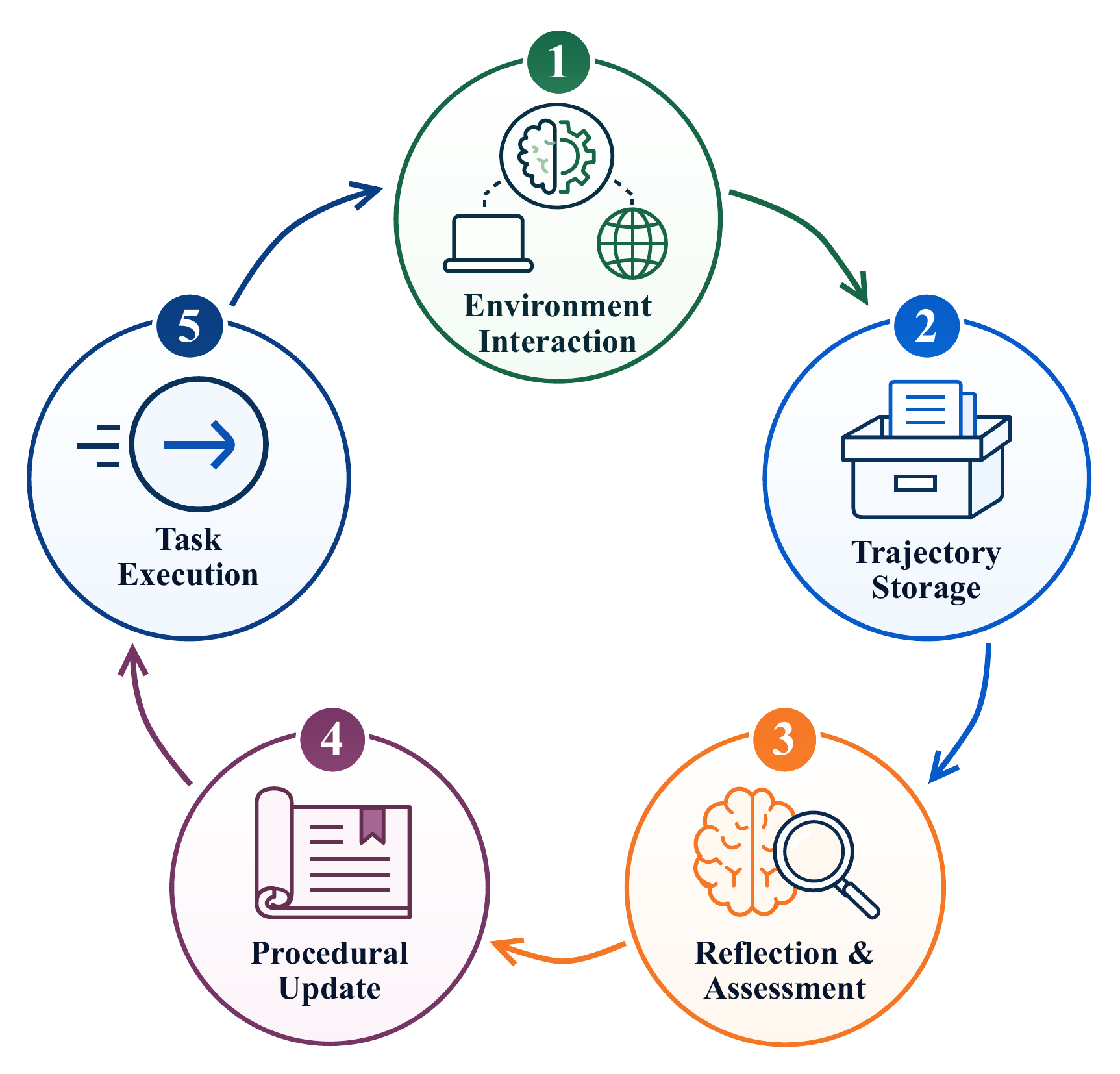}
\caption{Self-evolving agent memory loop. Agent memory captures interaction trajectories, retrieves them for reflection and assessment, consolidates insights into procedural or skill memory, and reuses the updated memory to guide future task execution in the environment.}
\label{fig:self_evolving_agent_memory_loop}
\end{figure}

Beyond storing information for recall, memory can support agent evolution. In this line of work, memory becomes a substrate for improving behavior: agents store experience, learn from failures, refine procedures, compress successful trajectories, and optimize the context supplied to the LLM. Figure~\ref{fig:self_evolving_agent_memory_loop} summarizes this self-evolving loop from environment interaction to trajectory storage, reflection, procedural update, and memory-guided task execution.

% Float placement remains flexible to avoid prematurely ending text pages.

\subsection{Reinforcement Learning and Continual Learning}\label{subsec:rl_continual_memory}

Experience-driven memory methods use previous interaction records as training or adaptation signals for future behavior. Their common goal is not merely to recall the past, but to convert past trajectories, errors, rewards, and reflections into better future action. The key design axis is how explicitly the system separates memory operations from task solving. \mth{MemAgent}{wang2024memagent} trains an agent to update internal memory while processing long contexts; \mth{Memory-R1}{zhang2025memoryr1} separates a Memory Manager with ADD, UPDATE, DELETE, and NOOP operations from an Answer Agent; and \mth{\MemAlpha}{wang2025memalpha} learns to maintain multi-component memories under downstream QA rewards. This separation matters because memory writing becomes an action with consequences, not a passive log.

Other systems emphasize deployment-time adaptation and self-evolution. \mth{Memento}{yao2025memento}, \mth{ReMe}{huang2025reme}, \mth{EvolveR}{zhao2025evolver}, \mth{MemRL}{zhang2026memrl}, \mth{ProcMEM}{zhang2025procmem}, \mth{CASCADE}{ma2025cascade}, and \mth{SEDM}{supp2025sedm} use accumulated experiences, cases, procedural abstractions, or distributed self-evolving memories to improve repeated tasks without always fine-tuning the backbone model. Recent work on contextual replay, skill optimization, fast-and-slow learning, long-horizon engineering agents, memory bottlenecks, neural garbage collection, predictive forgetting, test-time learning, nested learning, and task-core memory~\citep{supp2025contextualReplay,supp2026skillopt,supp2026fastSlow,supp2026mia,supp2026pask,supp2026mlEngineering,supp2026openclawRl,supp2026memtier,supp2026neuralGarbage,supp2026predictiveForgetting,supp2025e2eTtt,supp2025mlMaster,supp2025memevolve,supp2025learningJob,supp2025goalSearch,supp2025generalAgenticMemory,supp2025decisionMemory,supp2025nestedLearning,supp2025lightsearcher,supp2025latentLearning,supp2025memgen,supp2025resum,supp2025sparseMemoryFinetuning,supp2025taskCoreMemory} broadens the same theme. The unresolved issue is credit assignment: a system must determine which remembered experience caused improvement, which merely correlated with success, and which should be forgotten before it hardens into a brittle strategy.

\begin{table}[!htbp]
\centering
\caption{Classification of memory for agent evolution.}
\label{tab:evolution_classification}
\begingroup
\footnotesize
\setlength{\tabcolsep}{4pt}
\renewcommand{\arraystretch}{1.08}
\begin{tabularx}{\textwidth}{P{0.18\textwidth} P{0.24\textwidth} P{0.29\textwidth} Y}
\toprule
\textbf{Branch} &
\textbf{Stored experience} &
\textbf{Evolution role} &
\textbf{Representative methods} \\
\midrule
RL and continual learning &
Trajectories, rewards, failures, reflections, task outcomes &
Use remembered outcomes to improve exploration, planning, self-correction, and policy behavior &
MemAgent, Memory-R1, Mem-$\alpha$, Memento, ReMe, MemRL \\
\midrule
Procedural and skill memory &
Skills, workflows, tool-use patterns, reasoning templates &
Abstract successful or failed traces into reusable procedures, workflows, and skills &
MemSkill, ReasoningBank, Memento-Skills, AgentEvolver, FLEX \\
\midrule
Context and harness engineering &
Context summaries, files, traces, code, tool protocols, external state &
Expose, fold, prune, or externalize runtime state so long workflows remain recoverable &
ACON, AgentFold, SWE-Pruner, LCM, MemCompiler \\
\bottomrule
\end{tabularx}
\endgroup
\end{table}
% Float placement remains flexible to avoid prematurely ending text pages.

\subsection{Procedural and Skill Memory}\label{subsec:procedural_skill_memory}

Procedural memory stores reusable action knowledge rather than factual content. \mth{Agent Workflow Memory}{zorazrw2024agentworkflowmemory}, \mth{MemSkill}{liu2025memskill}, \mth{ReasoningBank}{ouyang2025reasoningbank}, \mth{Memento-Skills}{zhou2025mementoskills}, \mth{AgentEvolver}{xing2025agentevolver}, \mth{FLEX}{jiang2025flex}, and \mth{MUSE-Autoskill}{wang2025muse} all convert successful or failed experience into workflows, memory-operation skills, reasoning templates, policies, or executable procedures. Compared with episodic memory, procedural memory is more compact and easier to reuse across tasks; compared with parametric adaptation, it can remain inspectable as an external artifact.

The risk is that procedures abstract away the evidence that justified them. A workflow that succeeded because of a benchmark-specific shortcut may be stored as a general skill; a reasoning template may preserve a flawed assumption; and a tool-use policy may become unsafe when the environment changes. Procedural-memory systems therefore need skill validation, versioning, retirement, and rollback, not only retrieval. Their evaluation should measure transfer and robustness across tasks rather than repeated success on near-duplicate environments.

\subsection{Context Engineering and Harness Engineering}\label{subsec:context_harness_memory}

A third line treats memory as part of the agent runtime. \mth{ACON}{zhang2025acon}, \mth{AgentFold}{liu2025agentfold}, \mth{SWE-Pruner}{wang2025swepruner}, \mth{LCM}{xu2025lcm}, the \mth{CL-bench repo}{zhang2025clbench}, and the \mth{Code as Agent Harness survey}{wang2025codeasagentharness} emphasize that memory is not only stored content but also the surrounding interface that exposes files, traces, summaries, tools, and execution state. Related work on grep-like harnesses, task-specific memory harnesses, agentic file systems, context engineering, and state-conditioned memory compilation for embodied agents~\citep{supp2026grepHarness,supp2026mHarness,supp2025agenticFileSystem,supp2025agenticContextEngineering,supp2026memcompiler} strengthens this point.

The main contribution of this branch is operational rather than mnemonic: it decides which intermediate artifacts remain available and which are hidden, pruned, folded, or externalized. This makes harness memory powerful for software agents, web agents, and long workflows, but also makes evaluation difficult because performance gains may come from better tool exposure rather than better memory construction. A useful benchmark should therefore isolate whether the harness improved memory persistence, retrieval, and update behavior, or merely supplied more convenient context to the base model.

\subsection{Comparative Discussion}\label{subsec:evolution_discussion}

Memory for agent evolution differs from recall-oriented memory because success is measured by behavioral improvement rather than only answer accuracy. RL and continual-learning systems use memory as a feedback substrate; procedural memory abstracts reusable actions; and context and harness engineering preserves the intermediate state needed for reliable execution. These approaches are complementary: an agent may use episodic memories to learn a skill, store the skill as procedural memory, and expose the skill through a harness that manages files, tools, and execution traces. They also differ in risk profile. Experience memories may overfit to spurious successes, procedural memories may preserve brittle or unsafe workflows, and harness-level memories may expose stale files or tool states as if they were reliable evidence. The most important comparison is therefore not which method stores more experience, but which method proves that stored experience causally improves later behavior while remaining reversible, auditable, and safe to retire.

\section{Benchmarks}\label{sec:benchmarks}

Evaluating memory-augmented LLMs requires benchmarks that test not only retrieval accuracy but also persistence, temporal reasoning, personalization, multimodal grounding, hallucination control, and long-horizon task execution. Following the organization used in this survey, we divide benchmarks into plain-text benchmarks, multimodal benchmarks, and simulation environments.

\subsection{Plain-Text Benchmarks}\label{subsec:plain_text_benchmarks}

Plain-text benchmarks evaluate memory through conversations, user profiles, documents, and assistant histories. \mth{LoCoMo}{maharana2024evaluatinglongtermconversationalmemory} and \mth{LoCoMo+}{zhang2025locomoplus} focus on long-term conversational and beyond-factual cognitive memory, while \mth{LongMemEval}{wu2025longmemevalbenchmarkingchatassistants} evaluates information extraction, multi-session reasoning, temporal reasoning, knowledge updates, and abstention in long-term chat-assistant histories. \mth{BEAM}{zhang2025beam}, \mth{PersonaMem}{chen2025personamem}, \mth{MemoryAgentBench}{li2025memoryagentbench}, \mth{LifelongAgentBench}{zhang2025lifelongagentbench}, \mth{Minerva}{minerva2025}, \mth{HaluMem}{wang2025halumem}, \mth{LTM-QA}{supp2025ltmqa}, and \mth{TENURE}{supp2026tenure} emphasize different aspects of memory evaluation, including personalization, lifelong interaction, hallucinated memory use, long-context QA memory, precision-aware retrieval, operation-level hallucination diagnosis, and reliable abstention. Adjacent long-context benchmarks such as \mth{\InfinityBench}{zhang2024inftybenchextendinglongcontext}, \mth{LongBench}{bai2024longbenchbilingualmultitaskbenchmark}, \mth{LongBench v2}{bai2025longbenchv2deeperunderstanding}, and \mth{NoLiMa}{li2025nolimabenchmark} are useful controls for separating persistent memory mechanisms from simply placing more evidence inside a longer prompt.

\begin{table}[!htbp]
\centering
\caption{Classification of memory evaluation resources. The table separates evaluation target, memory capability, interaction format, and representative benchmark instead of using method-centered columns.}
\label{tab:evaluation_classification}
\footnotesize
\setlength{\tabcolsep}{3pt}
\begin{tabularx}{\textwidth}{P{0.14\textwidth} P{0.17\textwidth} P{0.19\textwidth} P{0.15\textwidth} Y}
\toprule
\textbf{Benchmark family} &
\textbf{Evaluation target} &
\textbf{Memory capability} &
\textbf{Task format} &
\textbf{Representative benchmarks} \\
\midrule
Plain-text benchmarks &
Conversation, user profile, long documents, assistant memory &
Multi-session recall, temporal update, abstention, hallucination control, personalization, long-context control &
QA, summarization, user modeling, memory judging &
\tcell{LoCoMo\\LoCoMo+\\LongMemEval\\BEAM\\PersonaMem\\MemoryAgentBench\\LifelongAgentBench\\Minerva\\HaluMem\\LTM-QA\\TENURE\\\InfinityBench\\LongBench\\LongBench v2\\NoLiMa} \\
\midrule
Multimodal benchmarks &
Video, image, egocentric stream, robotic memory &
Visual grounding, long video recall, temporal event tracking, multimodal personalization &
Video QA, image search, embodied memory tasks &
\tcell{Video-MME\\LVBench\\TeleEgo\\DeepImageSearch\\Persona-MME\\RoboMemArena} \\
\midrule
Simulation environments &
Apps, tools, feedback loops, evolving environments &
Long-horizon state tracking, task recovery, tool-use memory, procedural reuse &
Interactive app tasks, office workflows, simulated conversations &
\tcell{AppWorld\\MemoryBench\\AMemGym\\ARE\\OdysseyBench} \\
\bottomrule
\end{tabularx}
\end{table}
% Float placement remains flexible to avoid prematurely ending text pages.

Additional plain-text and text-adjacent benchmarks fill gaps around role-playing histories and memory-specific diagnosis. \mth{MOOM}{supp2025moom} evaluates maintenance, organization, and optimization in ultra-long role-playing dialogues; \mth{MemBench}{supp2025membench} broadens agent-memory evaluation across memory abilities; and \mth{StoryBench}{supp2023storybench} is useful as a text-to-video story-continuity benchmark for checking whether generated visual sequences remain consistent across temporally ordered prompts.

\subsection{Multimodal Benchmarks}\label{subsec:multimodal_benchmarks}

Multimodal benchmarks extend memory evaluation beyond text. \mth{Video-MME}{fu2024videomme} and \mth{LVBench}{wang2024lvbench} test long-video understanding and temporal reasoning. \mth{TeleEgo}{lin2025telego} evaluates egocentric multimodal memory, while \mth{DeepImageSearch}{zhang2025deepimagesearch} focuses on memory-guided image search. \mth{Persona-MME}{li2025personamme} and \mth{RoboMemArena}{chen2025robomemarena} further evaluate personalization and robotic or embodied memory. These benchmarks are important because caption-only memory can appear successful on text questions while failing to preserve object identity, spatial relations, or fine-grained perceptual evidence.

The multimodal benchmark landscape also includes long-video and sparse-memory QA resources that are adjacent to agent memory. \mth{MovieChat+}{supp2024moviechatplus} tests question-aware sparse memory for long video QA, \mth{CinePile}{supp2024cinepile} provides movie-scale QA, \mth{LongVideoBench}{supp2024longvideobench} evaluates interleaved long-context video-language understanding, and \mth{EgoSchema}{supp2023egoschema} diagnoses very long-form egocentric video understanding. These resources help separate genuine multimodal memory from short-window visual recognition.

\subsection{Simulation Environments}\label{subsec:simulation_envs}

Simulation environments evaluate memory through interactive task completion rather than static question answering. \mth{AppWorld}{trivedi2024appworld} builds a controllable world of everyday applications and APIs, requiring agents to remember user instructions, app states, and previous tool results during multi-step tasks. \mth{MemoryBench}{ma2024memorybench} evaluates whether agents can maintain and reuse memory under feedback-driven interactions. \mth{AMemGym}{liu2025amemgym} provides an interactive setting for testing conversational memory over long-horizon assistant sessions. \mth{ARE}{huang2025are} evaluates agents in adaptive or evolving environments where memory must support recovery and continued progress. \mth{OdysseyBench}{wang2025odysseybenchevaluatingllmagents} focuses on long-horizon office workflows, where agents must coordinate information across documents, spreadsheets, emails, calendars, PDFs, and intermediate files.

\subsection{Evaluation Metrics and Failure Modes}\label{subsec:evaluation_metrics_failure_modes}

A benchmark should reveal not only whether the correct answer is produced, but also whether memory was written, updated, retrieved, and used correctly. Table~\ref{tab:metrics_failure_modes} summarizes representative capabilities and failure modes that should be measured across memory systems.

\begin{table}[!htbp]
\centering
\caption{Capability-oriented evaluation dimensions for memory-augmented LLMs.}
\label{tab:metrics_failure_modes}
\footnotesize
\begin{tabularx}{\textwidth}{P{0.18\textwidth} P{0.24\textwidth} P{0.24\textwidth} Y}
\toprule
\textbf{Capability} & \textbf{What to measure} & \textbf{Typical failure mode} & \textbf{Possible metrics} \\
\midrule
Persistence & Whether important information remains available across sessions & The agent forgets durable facts or long-term preferences & Memory recall, answer accuracy over delayed queries \\
\midrule
Update and conflict handling & Whether newer evidence revises older memory & The agent retrieves stale or contradictory facts & Update accuracy, contradiction resolution rate \\
\midrule
Selective writing & Whether the system stores useful information without accumulating noise & Over-writing irrelevant details or under-writing important facts & Memory precision, memory recall, storage growth rate \\
\midrule
Retrieval grounding & Whether retrieved memory is relevant and correctly used & Irrelevant memory distracts reasoning or produces false continuity & Retrieval precision, citation/evidence correctness, downstream utility \\
\midrule
Abstention and hallucination control & Whether the model avoids inventing memories & The model claims to remember interactions that never occurred & False-memory rate, abstention accuracy \\
\midrule
Privacy and deletion & Whether sensitive memory is protected and removable & Private memory leaks across users or persists after deletion & Leakage rate, deletion success rate, access-control violation rate \\
\bottomrule
\end{tabularx}
\end{table}
% Float placement remains flexible to avoid prematurely ending text pages.

\section{Challenges and Future Directions}\label{sec:challenges}

\subsection{Unified Memory Organization}\label{subsec:unified_memory_org}

Current systems often combine text chunks, embeddings, graphs, KV caches, tool traces, and user profiles in ad hoc ways. Future systems need unified schemas that can align episodic, semantic, procedural, multimodal, and parametric memories without losing provenance or update history. A promising direction is to separate memory content from memory metadata: content stores what is remembered, while metadata records source, timestamp, confidence, permissions, access history, and deletion status. Such organization would make memory systems easier to audit, compare, and integrate across applications.

\subsection{Reliable Memory Writing}\label{subsec:reliable_writing}

Long-term systems fail when they write too much, write too little, or write incorrect abstractions. Better memory construction should distinguish durable facts, transient preferences, temporary task states, and uncertain observations. It should also support explicit contradiction handling and confidence-aware revision. Future research should move beyond heuristic summarization toward write policies that are evaluated against delayed downstream tasks, not only immediate extraction quality.

\subsection{Retrieval Beyond Similarity}\label{subsec:retrieval_beyond_similarity}

Nearest-neighbor similarity is insufficient for long-term memory. Useful memories may be old, indirect, private, contradicted, or only relevant under a procedural goal. Retrieval should combine semantic similarity with temporal reasoning, graph traversal, task intent, permission constraints, and uncertainty estimation. An important direction is multi-stage retrieval: first identify the task and memory type needed, then retrieve candidate memories, then verify recency, compatibility, and source reliability before injecting them into the model context.

\subsection{Multimodal Grounding and Video Intelligence}\label{subsec:multimodal_video_intelligence}

Multimodal memory must preserve object identity, spatial relations, audiovisual timing, and links to original perceptual evidence. Caption-only memory is compact but can discard important visual details. Future work should balance compact textual summaries with retrievable perceptual anchors such as frames, regions, audio segments, trajectories, or screen states. Evaluation should therefore test whether the system can recover evidence from the original modality rather than only answer questions from textual summaries.

Video-centered intelligence exposes an even broader infrastructure gap. Long-video understanding requires persistent temporal indexes that can track events, entities, causal dependencies, and scene changes across hours of perceptual input, while long-video generation requires memory mechanisms that maintain identity, layout, motion, and narrative consistency across extended outputs. World models and embodied agents further require memory substrates that connect perceptual evidence with latent scene state, physical constraints, action effects, and interaction history. Current systems still often depend on sparse frame sampling, caption-based compression, or short-lived context buffers, which weakens causal tracking and makes generated or embodied behavior drift over long horizons. Future research should therefore develop scalable video memory infrastructure that combines online compression, hierarchical temporal event graphs, multimodal retrieval over raw evidence and latent world states, and evaluation protocols spanning understanding, generation, world modeling, and embodied control.

\subsection{Governance, Privacy, and Deletion}\label{subsec:governance_privacy_deletion}

Long-term memory often contains personal, proprietary, or task-sensitive information. Multi-user, multi-agent, and edge--cloud settings require access control, provenance tracking, permission-aware retrieval, conflict management, secure deletion, and privacy-preserving personalized memory management~\citep{supp2026memprivacy}. Parametric memory is especially challenging because a single remembered fact may not be directly inspectable or removable. Practical memory systems should therefore expose deletion logs, memory scopes, user consent controls, and policy-aware retrieval filters. These governance mechanisms should be treated as core technical components rather than deployment afterthoughts.

\subsection{Failure Amplification and Misaligned Adaptation}\label{subsec:failure_amplification}

Memory can amplify errors because stored information influences future actions. A hallucinated fact that is written into memory can become a persistent false belief; an incorrectly extracted graph edge can redirect later multi-hop reasoning; and a noisy user profile can bias personalization across sessions. Agent-evolution memory introduces an additional risk: a workflow that succeeds for the wrong reason may be converted into a reusable skill, causing the agent to repeat brittle strategies or exploit benchmark-specific shortcuts. Future systems should therefore treat memory writing as a high-impact operation rather than a passive logging step. Useful safeguards include uncertainty-aware writes, source-linked memories, conflict detection, periodic memory review, and controlled mechanisms for skill retirement or rollback.

\subsection{Evaluation Reliability}\label{subsec:evaluation_reliability}

Current benchmarks cover long-context retrieval, conversational memory, multimodal understanding, and simulation environments, but evaluation remains fragmented. Future benchmarks should measure memory write quality, update correctness, abstention, hallucinated memory use, privacy leakage, temporal consistency, and downstream task success. They should also distinguish between genuine memory use and success through long-context pattern matching. One practical approach is to evaluate agents under controlled memory ablations, where memory contents are removed, corrupted, delayed, or replaced to test whether the model truly depends on the intended memory mechanism.

\section{Conclusion}\label{sec:conclusion}

Memory research for LLMs has moved from simple retrieval augmentation toward a diverse ecosystem of nonparametric stores, graph memories, multimodal memories, parametric memory modules, and self-evolving agent systems. This survey organized the literature around a lightweight cognitive-science-inspired conceptual motivation layer, three computational mechanism families---nonparametric memory, parametric memory, and memory for agent evolution---and a cross-cutting evaluation layer that tests whether these mechanisms produce durable benefits. Across these families, the central design questions are how to construct memory, how to manage and consolidate it, how to retrieve it reliably, how to forget or delete it safely, and how to evaluate its contribution to long-term behavior. The next generation of memory-augmented agents will likely combine text, graph, multimodal, activation, and procedural memories under explicit governance mechanisms, using cognitive concepts mainly as design vocabulary, moving LLMs from reactive generation toward persistent, adaptive, and trustworthy long-term collaboration.

\section*{Declarations}

\subsection*{Funding}
No specific funding was received for this work.

\subsection*{Conflicts of interest/Competing interests}
The authors declare that they have no known competing financial interests or personal relationships that could have appeared to influence the work reported in this manuscript.

\subsection*{Ethics approval}
Not applicable. This article is a survey and does not involve human participants, human data, animals, or clinical materials.

\subsection*{Consent to participate}
Not applicable.

\subsection*{Consent for publication}
Not applicable.

\subsection*{Data availability}
This survey is based on publicly available publications, benchmark descriptions, benchmark repositories, system documentation, and project pages. No new dataset was created or analyzed for this manuscript.

\subsection*{Code availability}
Not applicable. This article is a survey and does not introduce new code or software.

\subsection*{Authors' contributions}
All authors contributed to the manuscript.

\subsection*{Acknowledgments}
The authors thank the creators of the open-source memory systems, benchmarks, and documentation resources that support research on memory-augmented large language model agents.

\bibliography{sn-bibliography}

\end{document}